\documentclass{PoS}

\newcommand{\ave}[1]{\left\langle{#1}\right\rangle}
\newcommand{\bea}{\begin{eqnarray}}
\newcommand{\eea}{\end{eqnarray}}

\newcommand{\Det}{{\rm Det}}
\newcommand{\Tr}{{\rm Tr}}
\newcommand{\gf}{\gamma_5}

\newcommand{\re}[1]{(\ref{#1})}

\title{QCD in Infrared Region and Spontaneous
Breaking of the Chiral Symmetry}

\ShortTitle{IR QCD and SBCS }

\author{\speaker{Mirzayusuf  Musakhanov}
\\
        National University of Uzbekistan\\
        E-mail: \email{musakhanov@gmail.com}}

\abstract{
The spontaneous breaking of chiral symmetry (SBCS)
 is one of the most important phenomena of hadron 
 physics. It defines the properties of all the light mesons 
   and baryons. The Chiral Perturbation Theory (ChPT)
 encodes QCD hadronic correlators at low-energy region 
 in the terms the low-energy constants (LEC) -- the expansion parameters on 
    light quark current masses $m$ and external momenta $p$.
   The LEC's can be extracted from the phenomenology or from QCD lattice calculations. 
On the other hand, QCD instanton vacuum/instanton liquid model 
provides a very natural nonperturbative explanation of the SBCS. It provides a consistent framework for description of the pions and thus 
 may be used for evaluation of the LEC. 
Our aim is to calculate the vacuum properties and the LEC's within instanton vacuum model and confront with phenomenology and lattice results.}

\FullConference{XXI International Baldin Seminar on High Energy Physics Problems,\\
		September 10-15, 2012\\
		JINR, Dubna, Russia}

\begin{document}
\section{Introduction}
\subsection{Nonperturbative QCD in Infrared Region}
QCD in the chiral limit is a good approximation to the real world and left and right-handed quarks are decoupled. But the hadrons has no chiral doublets which means that the QCD vacuum breaks the chiral symmetry. One of the signals of the Spontaneous Breaking of the Chiral Symmetry (SBCS) is the presence of the nonzero chiral quark condensate, $\langle \bar qq\rangle\ne 0$. The understanding of SBCS is provided by Casher-Banks formula\footnote{Assumed thermodynamic limit means that the volume $V$ goes to infinity faster than quark mass $m$ goes
to zero.}:
\begin{equation}
\langle \bar qq\rangle=-\frac{\pi}{V}\nu(\lambda=0)\,,
\end{equation}
The chiral condensate is thus proportional to the averaged spectral density of the QCD Dirac operator $(i\hat\partial + g \hat A)$ at zero eigenvalues $\nu(\lambda=0)$~\cite{Schafer:1996wv,Diakonov:2002fq}.
The key moment is that the Dirac operator in the background of topologically nontrivial field may
has an exact zero modes with $\lambda=0$ and in an accordance with a general
Atiah--Singer index theorem the number of these modes equal to Pontryagin index or the
topological charge of the background field. Then, the SBCS
is due to of a delocalization of the would-be zero modes, induced by the
background of a vacuum topologically nontrivial fields, resulting from quarks hopping between them. On the other hand the QCD sum rules phenomenology requests that the QCD vacuum
has the gluon condensate \cite{SVZ}:  
\begin{equation}
\frac{1}{32\pi^2}\langle G_{\mu\nu}^a G_{\mu\nu}^a\rangle \simeq (200\; MeV)^4 . 
\label{glcond}
\end{equation}
The simplest way to explain both phenomena -- the chiral quark condensate and the gluon condensate
is the instanton vacuum/instanton liquid model (see reviews~\cite{Schafer:1996wv,Diakonov:2002fq}). In the model the gluon condensate is directly related to the instanton density as $\frac{1}{32\pi^2}\langle G_{\mu\nu}^a G_{\mu\nu}^a\rangle=N/V=\bar R^{-4} $, where from Eq.(\ref{glcond}) the average inter-instanton distance in Euclidian space $\bar R\simeq (200\,MeV)^{-1}=1\,fm$.
With this value for $\bar R$ and phenomenological value for the quark condensate 
$\langle \bar qq\rangle\simeq - (250\, MeV)^3$ the average size of the instanton is estimated as $\bar\rho\simeq 0.33\,fm$.

The main problem of the instanton vacuum model is the lack of the confinement. A generalizations of the model providing the confinement was proposed recently with the price of the introducing of the other topological objects.  Among them are calorons with non-trivial holonomy, BPS monopoles or dyons, vortices, etc..

\subsection{Lattice QCD in Infrared Region} 

In the recent decades QCD vacuum have been intensively studied by direct numerical simulations
on the lattice. The presence of non-trivial topological objects was demonstrated by using various configuration-smoothing methods. The typical view of the distributions of the action and topological charge density before and after few steps of the ``cooling'' was the following: 
before  ``cooling'' they are heavily dominated by perturbative zero-point fluctuations
of the gluon fields. The ``cooling'' suppresses these fluctuations which leads to a smooth background
coinciding with an ensemble of instantons and antiinstantons with $\bar\rho\simeq 0.36\,fm$ and $\bar R\simeq 0.89\,fm$~\cite{CGHN}.

But recently 
the QCD instanton vacuum model was challenged by the detailed lattice studies of topological charge density distribution by means of overlap Dirac operator (possess an exact chiral symmetry on the lattice) which shows a three-dimensional, laminar and highly singular structure which seems contradict the instanton and similar pictures~\cite{Horvath03,Ilgenfritzetal2007}. Probably the structure of the QCD vacuum much more rich and we have to add another topological objects mentioned above. 

\begin{figure}[h]
\centerline{\includegraphics[scale=0.35]{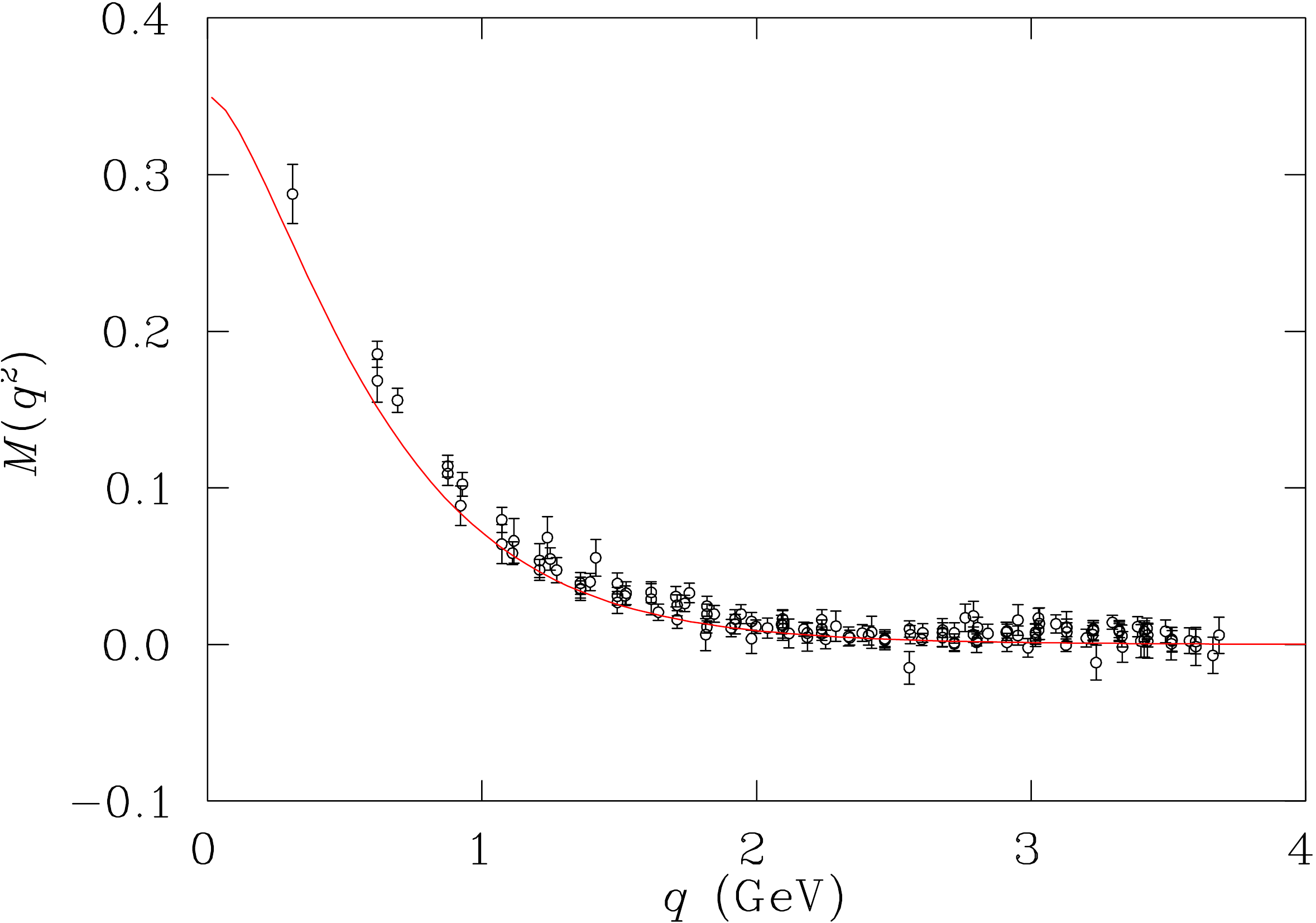}}
\caption
{Momentum dependence of dynamical quark mass $M(q)$ in the chiral limit. Points: lattice result ~\cite{Bowman:2004xi}. Red line: instanton vacuum model~\cite{Diakonov:1985eg}, {\bf no fitting}. }
\label{fig:Mlat}
\end{figure}

The question: which component of QCD vacuum is most important for the SBCS?
The answer is given by the comparison of the momentum dependence of the dynamical quark mass $M(q)$ in the chiral limit given at instanton vacuum model and lattice QCD, Fig.\ref{fig:Mlat}. Model $q$-dependence of $M(q)$ is mostly due to the quark zero-mode function in the (anti)instanton field. We see very good coincidence {\bf without any fitting} of the lattice results with the model. 

\begin{figure}[h]
\centerline{\includegraphics[width=9cm]{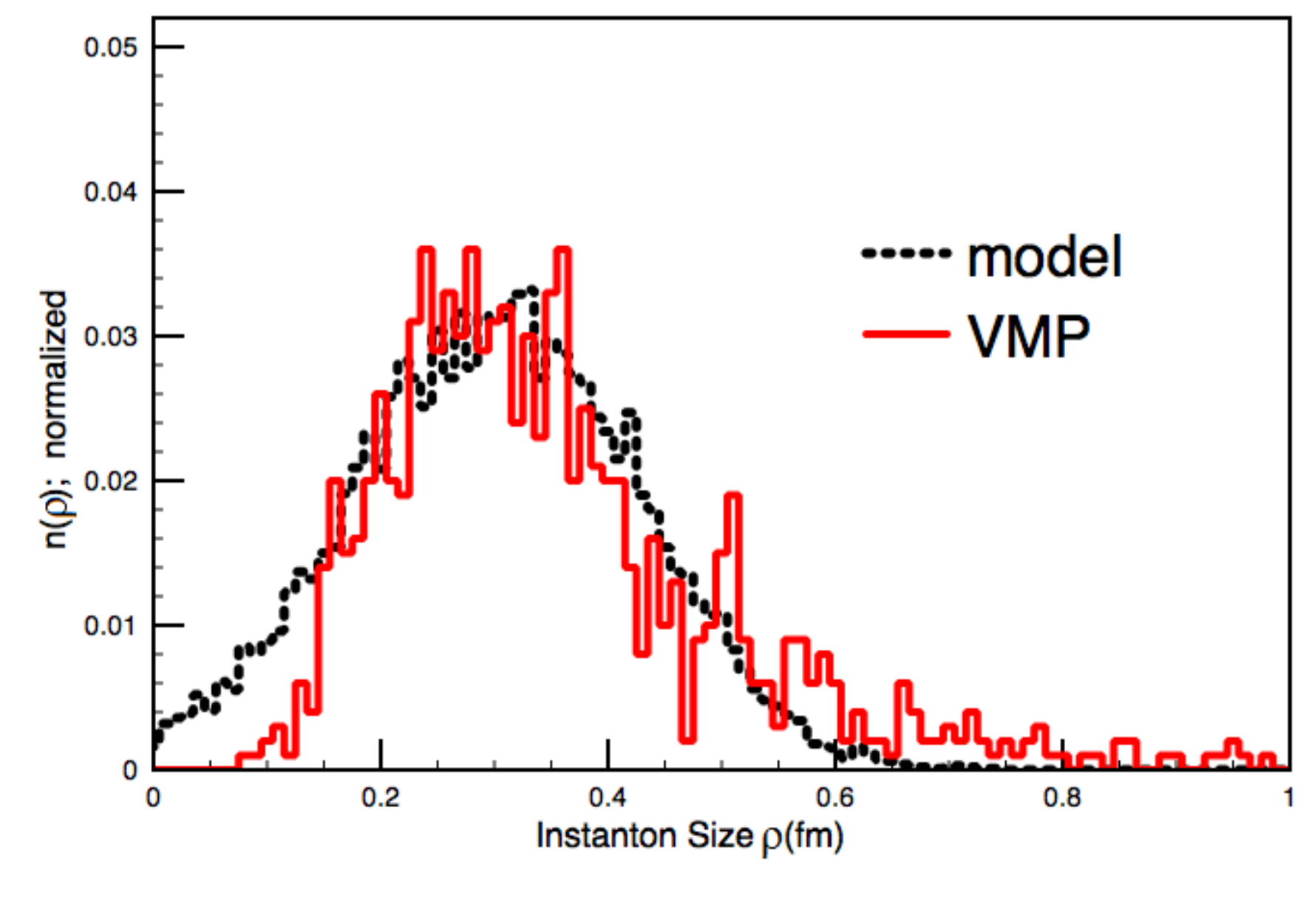}}
\caption{$n_{ model}(\rho) \propto \exp\left[ - \frac{(\rho-\bar\rho)^2}{2 \sigma^2}\right],$
with  $\bar\rho = 0.3~\textrm{fm}$, $\sigma = 0.13~\textrm{fm}$, $\bar R\simeq 1.07~\textrm{fm}$ 
and the VMP distribution from the corresponding 
lattice configurations~\cite{MilloFaccioli2011}. }
\label{fig:rholat}
\end{figure}

 The details of the instanton size distribution extracted from lattice simulations of the instanton liquid are presented at the Fig.\ref{fig:rholat}.

\subsection{QCD vacuum and pion physics} 

The properties of the QCD vacuum and its lowest excitations--pions are encoded in Low-Energy Constants (LEC) of Chiral Perturbation Theory (ChPT). It is natural to confront LEC's derived 
from the model with the phenomenology and lattice calculations of them.
ChPT gives the correlators in terms of LEC's as coefficients in the expansion over the current quark mass $m$ and external momentum of the order of $M_\pi$  (taking into account that in 
the lowest order $m_\pi^2\sim m$). This expansion have to take into account a chiral log terms $\sim \log m_\pi^2$ due to the contribution of the pion loops.
$N_f=2$ lowest $q^2$ order effective lagrangian $L_2$ have two  LEC's $F$ and $B$.  They correspond to the observables: 
 $F$ -- the pion decay constant in the chiral limit and $m_\pi^2= 2 B\, m$ -- the pion mass square in the lowest order on $m$~\cite{Gasser:1983yg}.
At $q^4$ order effective lagrangian $L_4$
has 10 independent \textit{bare} LEC's  $l_i, h_i$.
They are renormalized by pion loops to $\bar l_i$.
Physical observables should be expressed in terms of  $\bar l_i$~\cite{Gasser:1983yg}. 

Our aim is to calculate the vacuum properties and the LEC's within instanton vacuum model and compare with phenomenology and lattice results.

\section{ Instanton vacuum model}
Instanton $A^I$ (antiinstanton $A^{\bar I}$) is a solution of Yang-Mills equations in Euclidian space with topological charge $=1(-1)$ and correspond to the tunneling process of the gluon fields.
Instanton collective coordinates are
$
4\; {\mbox (centre)}\;\;+\;\;1\; {\mbox (size)}\;\;+\;\;
(4N_c-5)\; {\mbox (orientations)}\;\;=\;\;4N_c.
$
It is assumed that vacuum background is given by $A=\sum_{I}A^I +\sum_{\bar I}A^{\bar I}$. The main parameters of the instanton vacuum are average interinstanton distance $\bar R$ and average size  $\bar \rho$.  
The estimates of them are the following:
lattice estimates: $\bar R\approx 0.89\,fm$, $\bar \rho\approx 0.36\,fm$~\cite{CGHN},
phenomenological estimates: $\bar R\approx 1\,fm$, $\bar \rho\approx 0.33\,fm$~\cite{Schafer:1996wv},
our estimate (corresponding ChPT  $F_{\pi,{m=0}}=88 MeV, \langle\bar qq\rangle_{m=0}=-(255 MeV)^3$): $\bar R\approx 0.76\,fm$, $\bar \rho\approx 0.32\,fm$, 
Thus within $10-15\%$ uncertainty different approaches give similar estimates.
We see that the packing parameter  $\pi^2 (\frac{\bar \rho}{\bar R})^4 \sim 0.1$ is small
and independent averaging over instanton positions and orientations is justified.

\subsection{Light quarks in the instanton  vacuum}
We have to calculate the correlators beyond the chiral limit to extract LEC's. Our starting point is the interpolation formula~\cite{Musakhanov:1998wp,Musakhanov:vu,
Kim:2005jc,Goeke:2007bj,Goeke:2007nc} for the quark propagator in the field of single instanton:
 \bea
&&S_i=S_{0} - S_{0}\hat p \frac{|\Phi_{0i}\rangle\langle\Phi_{0i}|}{\langle\Phi_{0i}|\hat p S_{0} \hat p |\Phi_{0i}\rangle} \hat p S_{0},\,\,\, 
S_0=\frac{1}{\hat p + im},\,\,\,(i\hat\partial + g \hat A_i )\Phi_{0i}=0
\label{interpol}
\eea
where $\Phi_{0i}$ is the quark zero-mode function. The advantage of the formula shown by the projection of the propagator to the zero-mode 
 \bea
&&S_i|\Phi_{0i}\rangle = \frac{1}{im}|\Phi_{0i}\rangle,\,\,\, \langle\Phi_{0i}|S_i
 =\langle\Phi_{0i}|\frac{1}{im}.
 \eea
 as it must be. Also, the calculations of the single-instanton quark effective action
reproduce exactly most important for our problems 
the $(m\rho)^2\ln m\rho$-term from~\cite{Lee2009}.

On the other hand the interpolation formula provide the summing up the re-scattering series for the full propagator (in the presence also of the flavor external fields $\hat V=s+p\gamma_5+\hat v+\hat a\gamma_5$): 
\begin{eqnarray}
&&\tilde S -\tilde S_{0} = -\tilde S_{0}\sum_{i,j}\hat p |\phi_{0i}\rangle
\left\langle\phi_{0i}\left|\left(\frac{1}{\hat p \tilde S_{0}\hat p}\right)\right|\phi_{0j}\right\rangle\langle\phi_{0j}|\hat p \tilde S_{0},
\\
&&\nonumber|\phi_0\rangle=\frac{1}{\hat p}L \hat p |\Phi_0\rangle,\,\,\, 
\tilde S_{0}=\frac{1}{\hat p +\hat V+im },\,\,\, 
L_i(x,z_i)={\rm P} \exp\left(i\int_{z_i}^x dy_\mu( v_\mu(y)+a_\mu(y)\gamma_5)\right).
\end{eqnarray}
Gauge connections $L_i$ provide flavor gauge-covariance of the propagator $\tilde S.$

Let's represent the quark determinant as $\Det=\Det_{high}\times\Det_{low}$~\cite{Diakonov:1985eg}. The high-frequencies part of the quark determinant $\Det_{high} $ can be calculated in single-instanton approximation, while  the low-frequencies part $\Det_{low}$ get the contribution from the whole set of the instantons according the formula:
\bea
\ln \tilde\Det_{low}=\Tr\int_m^{\bar M} dm'\, \tilde S(m') =\ln \det \langle\phi_{0,i}|\hat p \tilde S_0^{fg}  \hat p|\phi_{0,j}\rangle,
\label{det}
\eea
The partition function for the  light quarks $Z_N[V]$ is given by the averaging of $\tilde\Det_{low}$ over instantons collective coordinates. It was done by means of the fermionization with constituent quarks $\psi^\dagger ,\psi $ and leads to the t'Hooft-like nonlocal quark interaction term $Y_{N_f}$ with $2N_f$ legs~\cite{Musakhanov:1998wp,Musakhanov:vu}.  In the following we consider the $N_f=2$ case, neglecting the influence of strange and heavier quarks for the pion physics observables.
  
Further exponentiation in $Z_N[V]$ introduce the integration over dynamical couplings $\lambda_\pm$ and
 we have the partition function in the form:
\begin{eqnarray}
\label{Z}
&& Z_N[V]=\int d\lambda_+d\lambda_-D\psi^\dagger D\psi e^{-S}, 
\\
&&\nonumber S=\psi^\dagger (i\hat \partial+\hat V+im)\psi +\sum_\pm\left(N_\pm \ln\frac{K}{\lambda_\pm}-N_\pm
+ \lambda _\pm Y_2^\pm\right),
\\
&&\nonumber Y_2^\pm=\int d\rho n(\rho)\left(\alpha^2 \det_f J^\pm+\beta^2 \det_f J^\pm_{\mu\nu}\right),\,\, \alpha^2=\frac{2N_c -1}{(N_{c}^{2}-1)2N_c},\,\,{\beta^2}=\frac{{\alpha^2}}{8N_c-4},
\\
&&\nonumber J^\pm_{fg}=(2\pi\rho)^2\psi^{\dagger}_f\bar L  F\frac{1\pm \gf}{2} F L\psi_g,\;\,\,\, J^\pm_{\mu\nu,fg}=(2\pi\rho)^2\psi^{\dagger}_f\bar L F \sigma_{\mu\nu}\frac{1\pm \gf}{2} F L\psi_g,\,\,
\bar L=\gamma_4 L^\dagger \gamma_4 .
\end{eqnarray}
The interaction term non-locality form-factor $F$ in the momentum space
is completely defined by Fourier-transform of the quark zero-mode function as:
\bea
F(k) = - \frac{d}{dt}[I_0 (t )K_0 (t ) 
- I_1 (t )K_1 (t )]_{t =\frac{|k|\rho}{2}}\approx\frac{1}{1+2(|k|\rho)^2},\, 
|k|\rho<3;\,\,\,\frac{\sqrt{2}}{(|k|\rho)^3},\, |k|\rho>3.
\eea
So, the range of the non-locality is given by the instanton size $\rho$, as was expected.

The main purpose of our work is the calculations of various correlators with account of the  ${\cal O}(1/N_c,\,m,\,m/N_c,\,m/N_c\,\ln m)$-corrections, which means double expansion over $1/N_c$ and $m$.
We estimated such NLO corrections~\cite{Kim:2005jc,Goeke:2007bj,Goeke:2007nc,Goeke:2010hm}:
\begin{enumerate}
\item The width of the instanton size distribution is ${\cal O}(1/N_c)$.We found that the finite width corrections are negligible, as were expected. For example, the account of the finite width parameter $\sigma$ (see $n_{ model}(\rho)$ from the Fig.\ref{fig:rholat}) lead to the corrections $\approx 2.6\%$ for $\ave{\bar qq}$ and $\approx 5\%$ for $F_\pi^2$.
\item The back-reaction of the light quark determinants to the instanton vacuum properties does not sizably change the distribution over $N_++N_-$ but radically change the distribution over $N_+ - N_-$. Any $m_f=0$ leads to $\delta$-function type of the latter distribution. In the following we take $N_+ = N_-$.
\item There are the quark-quark tensor interaction terms which are $1/N_c$-suppressed and thus are absent in the LO. It might give the contribution to the vacuum magnetic susceptibility $\chi$. Our estimate shows that it leads to the correction $\le 1\%$ to the $\chi$, which is certainly negligible. 
\item We found the importance of the meson loops contribution. And most important among them are pion loops, certainly. They lead to the chiral logs and numerically large corrections.
\end{enumerate}
Though the evaluation of the meson loop corrections in the instanton vacuum model is similar to the earlier meson loop evaluations in the NJL model \cite{Nikolov:1996jj,Plant:2000ty,Pena:1999sp,Oertel:2000cw} important differences should be mentioned:
\begin{enumerate}
\item  
 Due to nonlocal formfactors there is no need to introduce independent fermion and boson cutoffs $\Lambda_f, \Lambda_b$. The natural cutoff scale for all the loops (including meson loops) is the inverse instanton size $\rho^{-1}$.
\item 
 In the instanton vacuum model the quark coupling constant $\lambda$ is defined through the saddle-point equation  whereas it is a fixed external  parameter in NJL.
\end{enumerate}
\section{Vacuum properties}
\begin{figure}[h]
\centerline{\includegraphics[scale=0.3]{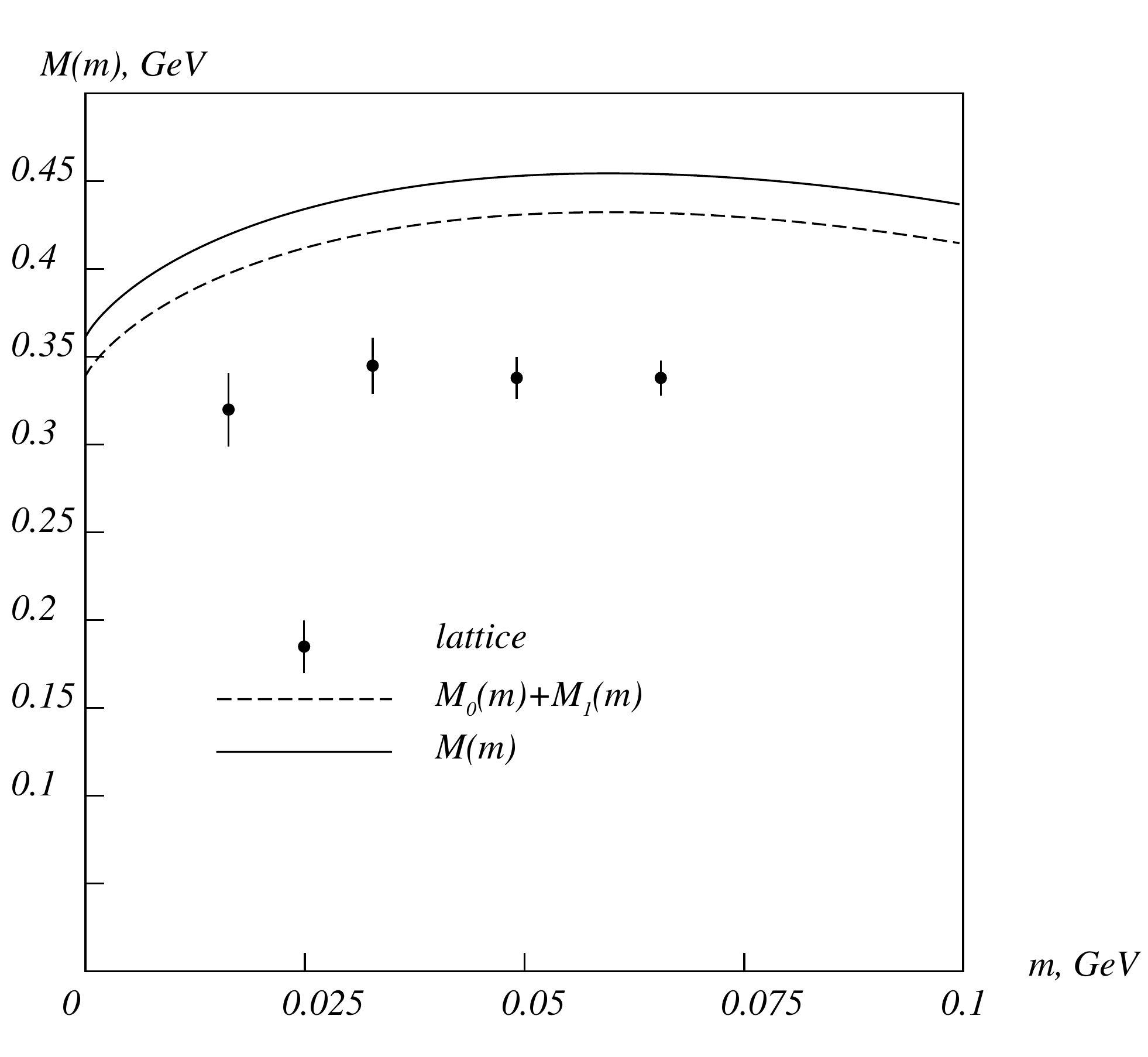}\includegraphics[scale=0.3]{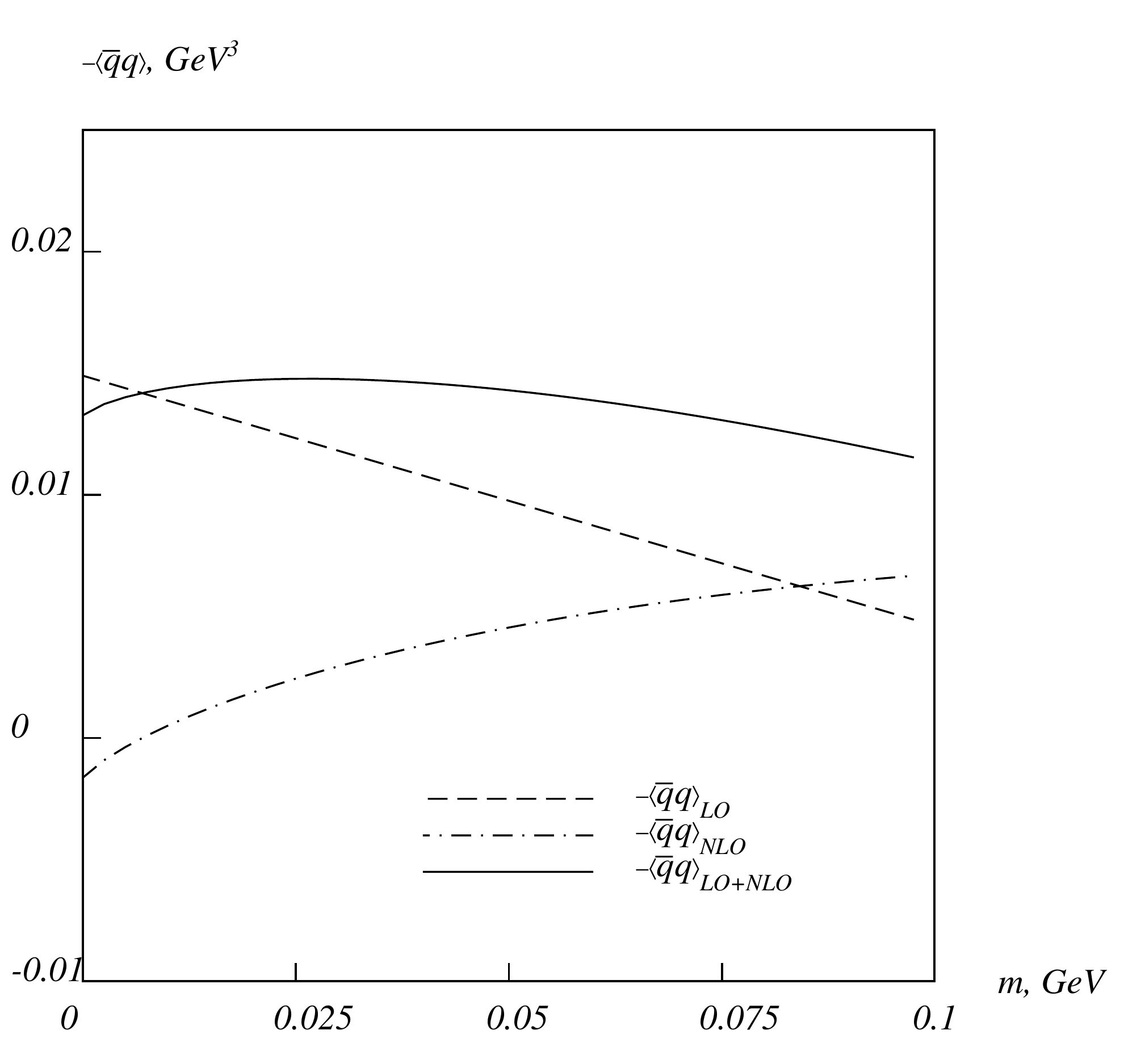}}
\caption[]{Left panel: $m$-dependence of the dynamical quark mass $M$ on the scale $\bar\rho^{-1}\approx 0.6\, GeV$.
The solid curve -- the exact numerical solution Eq.~\re{gap1_solution_exact}.
The dashed curve -- the solution obtained by the $1/N_c$ iterations with the same accuracy. 
 Lattice data points are from~\cite{Bowman:2005vx}, where the scale is $1.64 \,GeV$. \\
 Right panel: $m$-dependence of the quark condensate $-\ave{\bar qq}$ Eq. \re{qq}.  
The long-dashed curve is the LO result, the short-dashed curve is the NLO contribution, 
the solid curve is the total one. 
The dot-dashed line is the leading-order in $1/N_c$-expansion.}
\label{Mqq}\end{figure}

First it were calculated the dependencies of the dynamical quark mass $M$ and the quark condensate $\ave{\bar qq}$ on current quark mass $m$~\cite{Kim:2005jc}, see Fig.~\ref{Mqq}. 

We found $M(m)$ (in $GeV$) and $\langle\bar qq\rangle (m)$ (in $GeV^3$)  as a functions of $m$ (in $GeV$) as
\bea
 \label{gap1_solution_exact}
M(m)=0.36-2.36\,m -\frac{m}{N_c}(0.808+4.197 \ln m)+{\cal O}\left(m^2,\frac{1}{N_c}\right),
\\
-\langle\bar qq\rangle (m) =\left( 0.00497 - 0.034 m \right) N_c
 +\left(0.00168 - 0.049 m
 - 0.058 m\,\ln m\right)
 +{\cal O}\left(m^2,\frac{1}{N_c^2}\right)
 \label{qq}
 \eea
We see a very essential contributions of the chiral logs to the $m$ dependencies of $M(m)$ and $\ave{\bar qq}(m)$. They are coming from the pion loops, certainly. 

\subsection{Vacuum magnetic susceptibility}

External electromagnetic field $F_{\mu \nu }$ generate the quark currents in the QCD vacuum, described by the magnetic susceptibility $\chi _{f}(m_f) $ (with normalization factor $\langle i\psi^\dagger \psi \rangle_0=-\langle\bar qq\rangle (m=0)$): 
\bea
\langle0|\psi_f^\dagger \sigma_{\mu\nu}\psi_f|0\rangle_F = e_f
\,\chi_f(m_f)\, \langle i\psi^\dagger \psi \rangle_0\,F_{\mu\nu} 
\label{chi}
\eea
QCD sum rules~\cite{Ioffe:1983ju,Belyaev:1984ic} gave
$\chi _{f}(m_f)\,\langle i\psi^\dagger \psi \rangle_0\sim 40-70\, MeV$, while recent lattice measurements gave for this quantity $\approx 46(3)\, MeV$ (quenched SU(2) QCD, chiral limit)~\cite{Polikarpov09},
$\approx 52\, MeV$ (quenched SU(3) QCD, chiral limit)~\cite{Polikarpov10}, for (u,d)-quarks 
$\approx 40(1.4)\, MeV$ and 
for s-quark $\approx 53(7.2)\,MeV$ (fully dynamical QCD)~\cite{Bali2012}.
On the other hand, the magnetic susceptibility is measurable in jet production at $q_\perp\gg \Lambda_{QCD}$ at the process $\gamma+N\to (\bar qq)+N$ with polarized photon~\cite{Braun2002}. 

Our result for $\chi(m)\, \langle i\psi^\dagger \psi \rangle_0$~\cite{Goeke:2007nc} with the accuracy $\mathcal{O}\left(m^2,\frac{1}{N_c^2}\right)$ is
\bea
\chi(m)\, \langle i\psi^\dagger \psi \rangle_0 = N_c\left(0.015 + 5\cdot 10^{-4}m + \frac{m}{2\pi^2}\ln m \right) 
-0.007 - 0.415 m - 0.198 m \ln m.
\label{chi1}
\eea
\begin{figure}[h]
\center\includegraphics[scale=0.35]{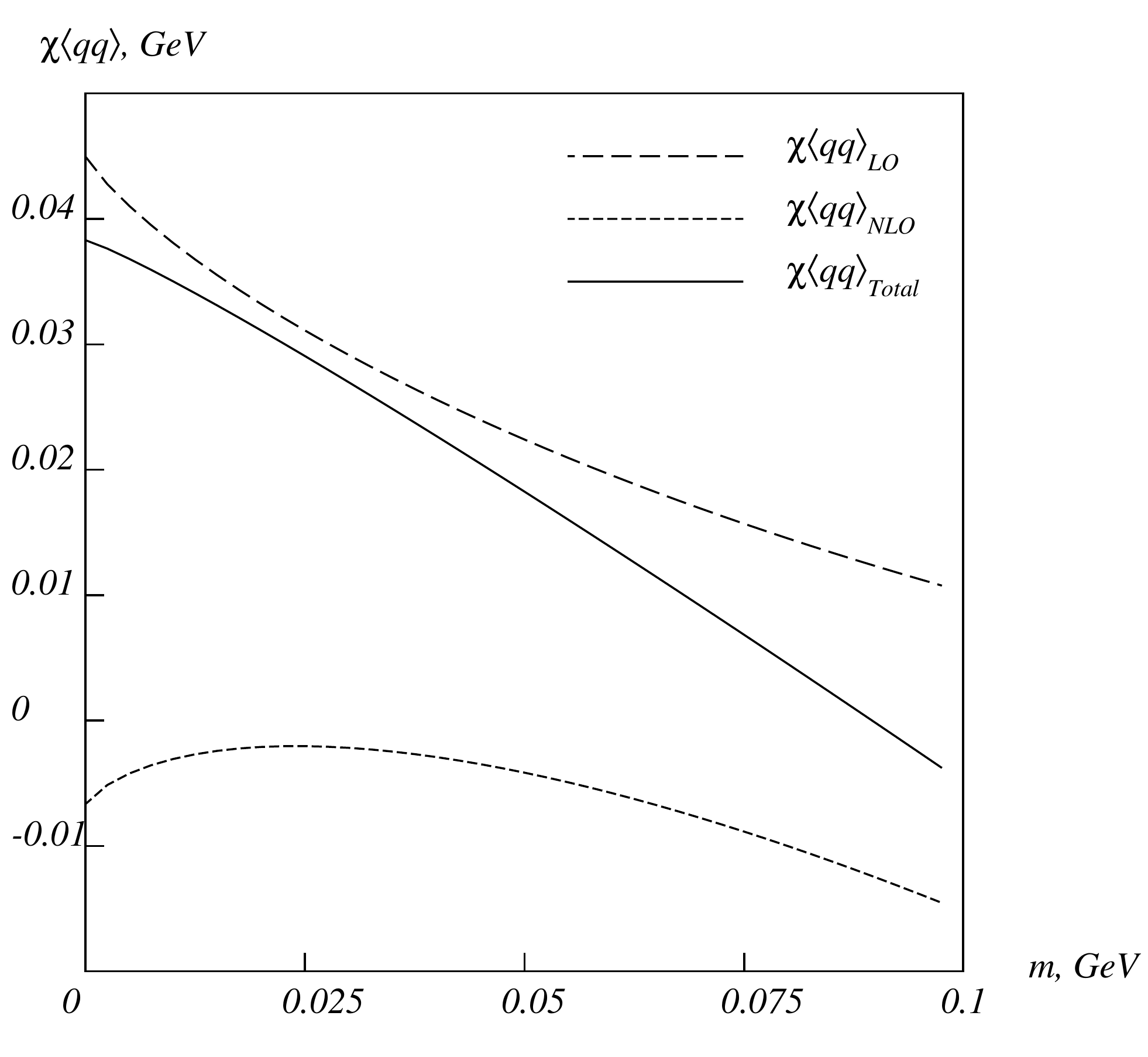}
\caption{Magnetic susceptibility $\chi(m)\langle i\psi^\dagger \psi \rangle_0$ as a function of current quark mass $m$.}
\label{fig.chi}
\end{figure}

Last term in~\re{chi1} is in the correspondence with the newly established chiral log theorem:
\bea
 \chi(m)=\chi(0)\left(1-\frac{3 m_\pi^2}{32\pi^2 F^2}\ln m_\pi^2\right)
\eea
We see that the strange quark contribution to the magnetic susceptibility is much less than lightest quarks one, Fig.~\ref{fig.chi}. This $m$-dependence contradict to the $m$-dependence obtained at~\cite{Bali2012} and the problem needs further research.

\section{Pion properties and low-energy constants   $\bar l_3$,  $\bar l_4$}

We calculated the two-point axial-isovector currents correlator~\cite{Goeke:2007nc}:
\begin{eqnarray}
\label{aa:structure}
\int d^4 x e^{-iq\cdot x}\langle j^{A,i}_\mu(x)j^{A,j}_\nu(0)\rangle
=\delta_{ij}F_\pi^2\left(\delta_{\mu\nu}-\frac{q_\mu q_\nu}{q^2+M_\pi^2}\right)+{\cal O} (q^2)
\end{eqnarray}
Here  $M_\pi$ has a meaning of pion mass and $F_\pi$ -- pion decay constant.

\begin{figure}[h]
\centerline{\includegraphics[scale=0.3]{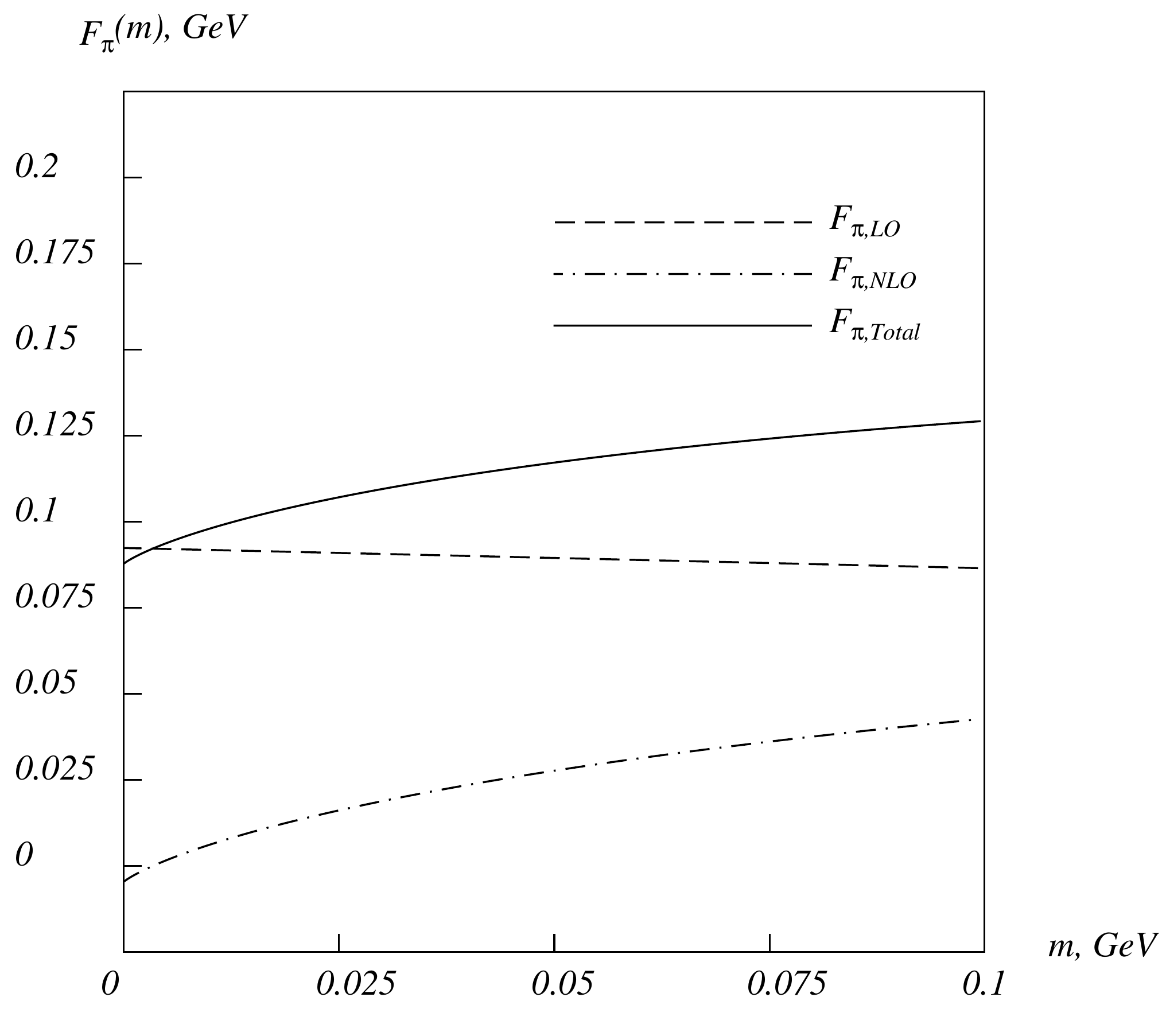}\includegraphics[scale=0.3]{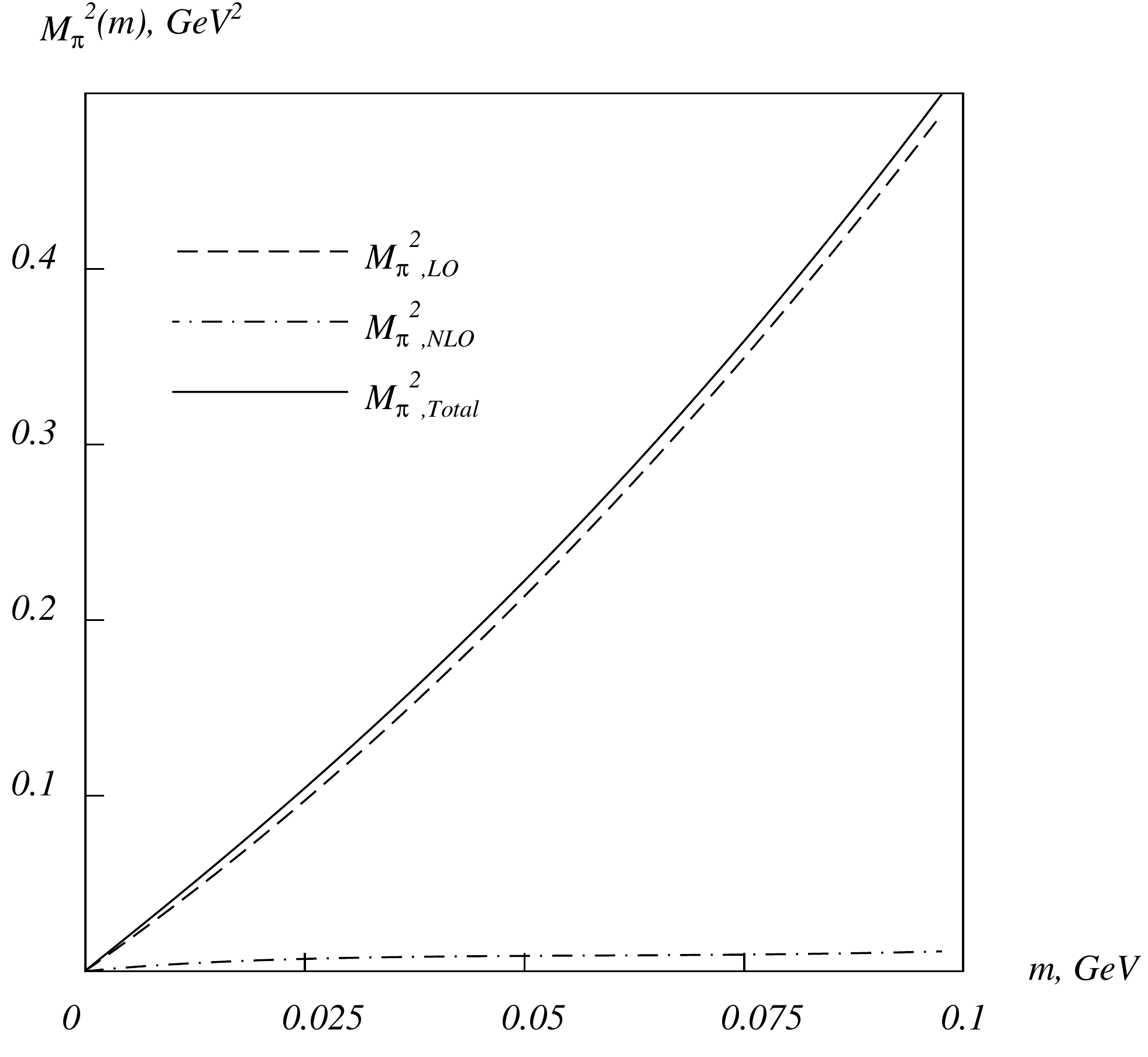}}
\caption[]{Left panel: $m$-dependence of the pion decay constant $F_\pi$.  The long-dashed curve is the LO contribution, the short-dashed curve is the NLO contribution, the solid curve is the total LO+NLO contribution.  The dot-dashed line represents the leading-order in $1/N_c$-expansion result, evaluated with the mass $M_0$. \\
Right panel: $m$-dependence of the pion mass  $M_\pi$.  The long-dashed curve is the LO contribution, the short-dashed curve is the NLO contribution, the solid curve is the total LO+NLO contribution. The dot-dashed line represents the leading-order in 
$1/N_c$-expansion result, evaluated with the mass $M_0$.}
\label{fig:FpiMpi}
\end{figure}

We found 
\begin{eqnarray}
&&\nonumber F_\pi^2=N_c\left(\left(2.85 -\frac{0.869}{N_c}\right)-\left(3.51+\frac{0.815}{N_c}\right)m-
\frac{44.25}{N_c}\,m\,\ln m +{\cal O} (m^2)\right)\cdot 10^{-3}\; [GeV^2]
\\
&& M_\pi^2=m\left(\left(3.49+\frac{1.63}{N_c}\right)+
m\left(15.5+\frac{18.25}{N_c}+\frac{13.5577}{N_c} \ln m \right)+{\cal O}(m^2)\right)[GeV^2]
\end{eqnarray}
represented by Fig.\ref{fig:FpiMpi}.

 According to~\cite{Gasser:1983yg}, the low-energy constants $\bar l_3$, $\bar l_4$ of the chiral lagrangian may be extracted from the ${\cal O}(m)$-corrections to physical quantities, e.g.
 \begin{eqnarray}
 M_\pi^2=m_\pi^2\left(1-\frac{m_\pi^2}{32\pi^2 F^2}\bar l_3+{\cal O}(m_\pi^4)\right),\,\,\,
 F_\pi^2=F^2\left(1+\frac{m_\pi^2}{8\pi^2 F^2}\bar l_4+{\cal O}(m_\pi^4)\right).
 \end{eqnarray}
Here  $m_\pi^2=2\,m\,B$. Lowest order LEC's $B=2.019\, GeV $ and $ F=88\, MeV$ were taken as an input to fix main instanton vacuum parameters as $\bar\rho=0.350\, fm$ and $\bar R=0.856\, fm$. 

\begin{figure}[h]
\centerline{\includegraphics[scale=0.25]{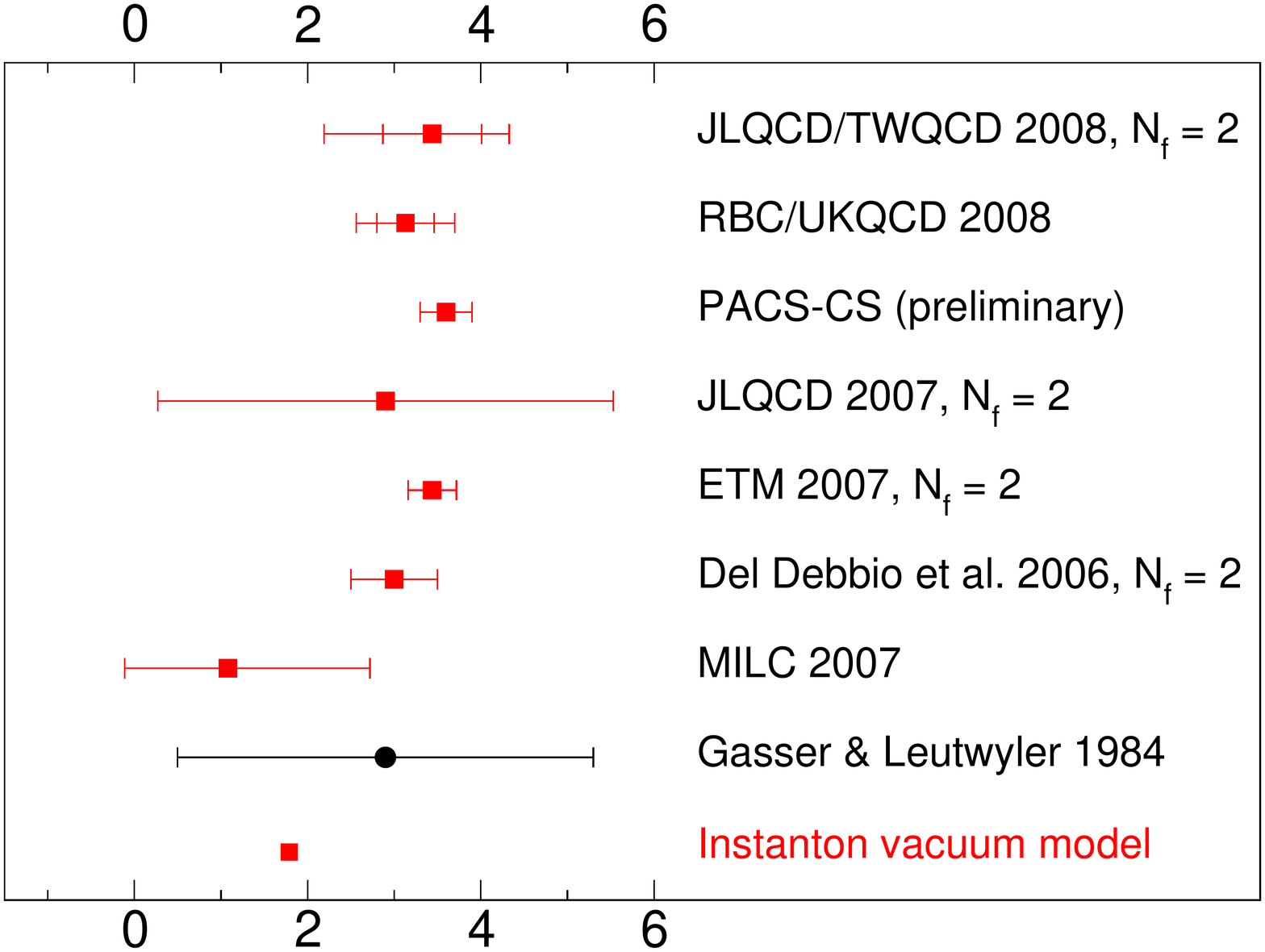}\includegraphics[scale=0.25]{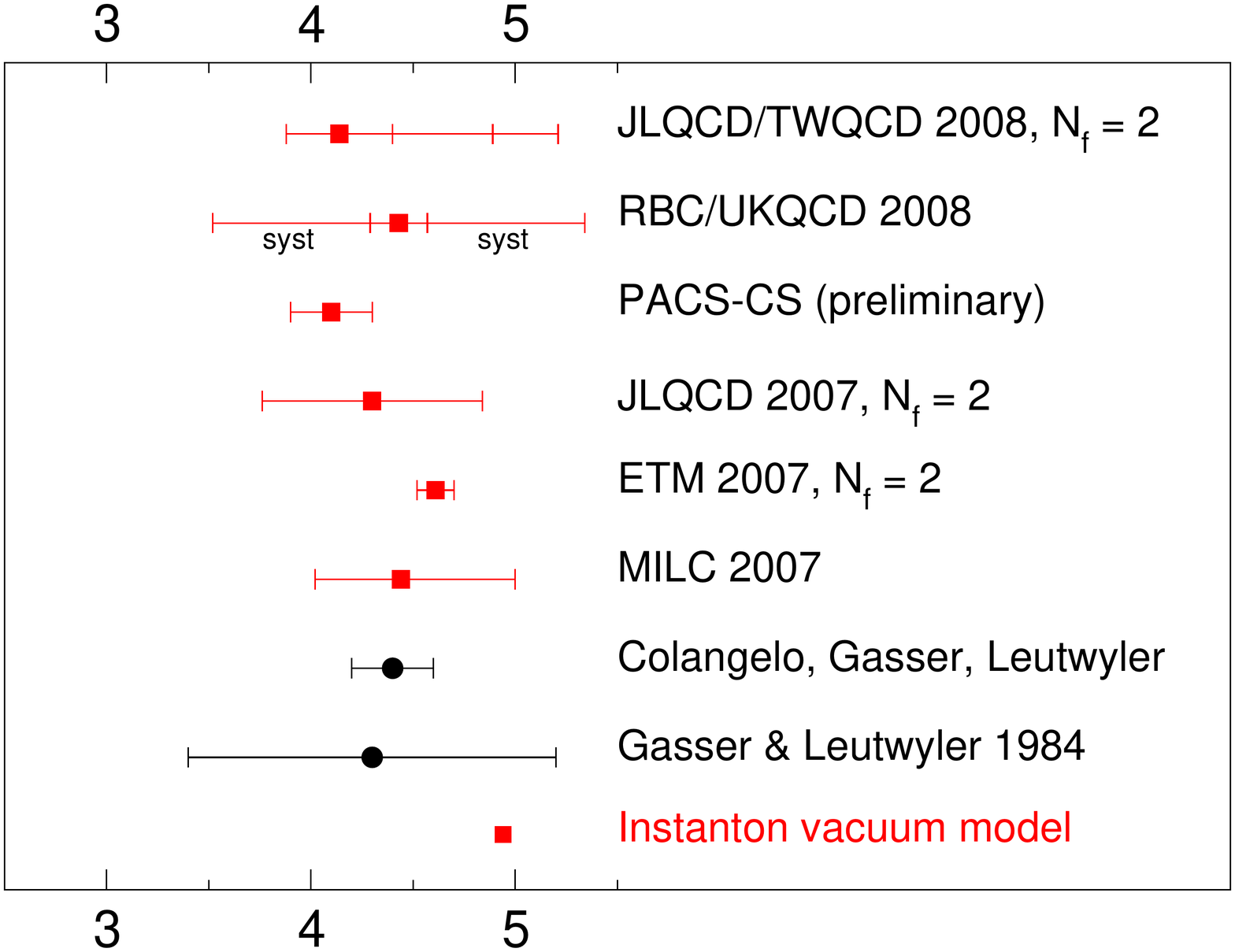}}
\caption{The LEC's  $\bar l_3$ (left panel) and $\bar l_4$ (right panel) -- 
recent lattice results from different collaborations~ \cite{JLQCD2008,MILC,Del Debbio,ETM,JLQCD,RBCUKQCD,PACS-CS}, phenomenological estimates~\cite{Gasser:1983yg,Colangelo} and our result.}
\label{l3l4}
\end{figure}

Then, we found (neglecting by the terms  ${\cal O}\left(\frac{1}{N_c}\right)$)
\begin{eqnarray}
 \bar l_3= - 1.1425 N_c+0.0738 - 0.999\ln m,\,\,\,
\bar l_4=- 0.0793N_c+0.01876 - \ln m,
 \end{eqnarray}
 which gives at $m=0.0055\,\,GeV$,
 $\bar l_3=1.84,\;\bar l_4=4.98$
 and  corresponding $M_\pi=0.142\,\,GeV$, $F_\pi=0.0937\,\,GeV$.
Our values of ($\bar l_3, \bar l_4$) should be compared with the phenomenological estimates~\cite{Gasser:1983yg,Colangelo} as well as lattice predictions~ \cite{JLQCD2008,MILC,Del Debbio,ETM,JLQCD,RBCUKQCD,PACS-CS} given in 
Fig.\ref{l3l4}.

\subsection{Isospin breaking by $m_u-m_d\not= 0$  and LEC's $h_3$ and $l_7$}
The correlator of the iso-vector pseudoscalar and iso-singlet pseudoscalar currents and
splitting between the $\ave{\bar{u}u}$  and $\ave{\bar{d}d}$ quark condensates in QCD 
are proportional to $m_u-m_d$ and described by LEC's $h_3$ and $l_7$ as
\bea
i\int dx\left\langle P^{3}(x)P^{0}(0)\right\rangle e^{iqx}=\frac{8B^{3}(m_{u}-m_{d})l_{7}}{q^{2}-M_{\pi}^{2}}+O(q^{2}),\,\,\,
\ave{\bar{u}u} -\ave{\bar{d}d}= 4B^{2}(m_u-m_d)h_3.
\eea
LEC $l_7$ define the pion masses shift due to $m_u-m_d\not= 0$:
\bea
(M^2_{\pi_0}-M^2_{\pi_\pm})_{\delta m}=-(m_u-m_d)^2\frac{2B^2}{F^2}l_7
\approx -(m_u-m_d)^2 1.2\cdot 10^{-3}l_7 .
\eea
Phenomenological estimate (GL, AP84) $l_7\sim 5\cdot 10^{-3}.$ 
Our estimates~\cite{Goeke:2010hm} for $l_7$ and $h_3$ are
\bea
l_{7}\approx(6.6\pm2.4)\cdot 10^{-4},\,\,\, h_3\approx5.48\cdot 10^{-3}.
\eea
We found strong dependence of the $l_7$ on the instanton vacuum parameters $\bar\rho$ and $\bar R$.
Confronting with the possible lattice calculation of this one will fix these parameters well.

\section{Summary}
SBCS is generated by the  `hopping' of quarks between topologically nontrivial gluon lumps leading to 
changing of quark chirality.
The strong evidence  in favor of the instanton vacuum model given by the momentum dependence 
of the dynamical quark mass $M(p)$ found in the lattice calculations (see Fig.\ref{fig:Mlat}).

The proposed interpolation formula~\re{interpol} for the quark propagator in the single instanton field and in the presence of the flavor external fields lead to the low-frequencies quark determinant~\re{det}. 

Next task was a derivation of the light quarks partition function $Z[V]$, which is a functional of the external flavor fields $V$. For this one we averaged the quark determinant over instantons, using diluteness of the instanton media. On this way it was introduced constituent quarks as a tool of this averaging. 

The partition function~\re{Z} has t'Hooft-like interaction terms with $2N_f$ quark legs, where the coupling $\lambda$ is not a fixed one, but must be calculated from saddle-point condition. We considered $N_f=2$-case. 

Since we are going beyond of the chiral limit we have to calculate the correlators with account of $O(1/N_c, m, m/N_c, m/N_c \ln m)$-corrections. Within the model there are the contributions of meson loops, finite width of instanton size distribution and quark-quark tensor interactions term to these corrections. We found that the most important are meson loops, while other contributions are negligible. 

ChPT describes SBCS in low-energy region in the terms of LEC's. We fixed the main parameters of the model $\bar R$ and $\bar\rho$ in terms of lowest $p^2$ order LEC's $B$ and $F$. With these parameters we also considered the vacuum magnetic susceptibility $\chi _{f}$ and its dependence on the current quark mass $m_f$, see Fig.~\ref{fig.chi}. It might be important for the phenomenology of the jet production at the process $\gamma+N\to (\bar qq)+N$ with polarized photon.  

At further step we calculated next $p^4$ order LEC's  
$\bar l_3$, $\bar l_4$, $l_7$ and $h_3$. Most important among them are $\bar l_3$ and $\bar l_4$, since there are available phenomenological and lattice estimates of them. We confronted our model calculation with available data and found reasonable correspondence, see Fig.~\ref{fig:FpiMpi}. 
  
This means that the instanton vacuum is applicable for understanding of the low-energy hadron physics, at least on the qualitative level.

I would like to thank Prof. Ernst-Michael Ilgenfritz for useful discussions during ISHEPP2012.

\end{document}